# Orientation-dependent etching of silicon by fluorine molecules: a quantum chemistry computational study

Omesh Dhar Dwivedi [1,2], Yuri Barsukov [1,a], Sierra Jubin[1,3], Joseph R. Vella[1], and Igor Kaganovich[1]

[1]Princeton Plasma Physics Laboratory, Princeton University, Princeton, NJ, United States of America, 08543
[2]Drexel University, Philadelphia, PA, United States of America, 19104
[3]Princeton University, Princeton, NJ, United States, 08544

[a)] Electronic mail: yvbarsuk@pppl.gov

Anisotropic etching is a widely used process in semiconductor manufacturing, in particular for micro- and nano-scale texturing of silicon surfaces for black silicon production. The typical process of plasma-assisted etching uses energetic ions to remove material in the vertical direction, creating anisotropic etch profiles. Plasma-less anisotropic etching, considered here, is a less common process that does not use ions and plasma. The anisotropy is caused by the unequal etching rates of different crystal planes; the etching process thus proceeds in a preferred direction. In this paper, we have performed quantum chemistry modeling of gas-surface reactions involved in the etching of silicon surfaces by molecular fluorine. The results confirm that orientation-dependent etch rates are the reason for anisotropy. The modeling of $F_2$ dissociative chemisorption on the F-terminated silicon surfaces show that Si-Si bond breaking is slow for Si(111) surface, while it is fast for the Si(100) and Si(110) surfaces. The Si(100) and Si(110) surfaces incorporate a larger number of fluorine atoms resulting in the Si-Si bonds having a larger amount of positive charge which lowers the reaction barrier of $F_2$ dissociative chemisorption, yielding a higher etch rate for the Si(100) and Si(110) surfaces compared to the Si(111) surfaces. Molecular dynamics modeling of the same reactions has shown that the chosen reactive bond order (REBO) potential does not accurately reproduce the lower reaction barriers for $F_2$ dissociative chemisorption on Si(100) and Si(100) surfaces. Thus, reparameterization is necessary to model the anisotropic etching process that occurs at lower temperatures.

## I.  INTRODUCTION

Orientation-dependent etching of silicon has been used for the texturing of silicon surfaces via dry and wet etching processes to produce black silicon[1,2]. The etch rate in the (111) direction is much slower than in the (100) and (110) directions, and the resulting anisotropic etching process can be used to create nanostructures. Plasma-less atmospheric dry etching (ADE) commercialized by Nines Photovoltaics is a cost-effective method of nano-scale silicon surface texturing. A schematic of this process is shown as Figure 1, illustrating the experimental etching results from[3] at 3.33% $F_2$ in air and substrate temperatures of 200 and 300°C. In the ADE process implemented by Nines Photovoltaics, a silicon wafer is transported on a heated conveyer through a reaction chamber containing the $F_2$ gas. The substrate temperature, $F_2$ concentration, and duration of the exposure to $F_2$ all control the etch process. Kafle et al. [3–6] showed that at temperatures higher than 200°C, a nano-scale pits are formed on an partly oxidized Si(100) surface during plasma-less etching. Nucleation of the pits due to anisotropic etching begins at reactive sites where the oxide layer is absent. The slope of the pit walls is defined by the angle between the Si(111) surface and Si(100) surface; these planes intersect at an angle of about 55 degrees, as shown in Fig. 1b and 1d. The cone-shaped nanostructures shown in Fig. 1b gradually disappear with increasing temperature (see Fig. 1c) as the etching process becomes more isotropic. It is schematically shown in Fig. 1d how the anisotropic etching profile appears on Si(100) surface due to slow etching of Si(111). Our focus is limited to analysis of anisotropic etching on a molecular level, using DFT (density functional theory) simulations of reaction pathways and investigating the suitability of a reactive bond order (REBO) molecular dynamics potential for simulating this anisotropic etching process. We did not study the effects of $F_2$ concentration or process duration on the etch anisotropy during ADE processing; nor have we examined the process of pit nucleation on the partly oxidized surface. It should be noted that a similar effect, where silicon is slowly etched in the (111) direction compared to the (100) direction by $Cl_2$/Ar inductively coupled plasma, was observed by Du et al. [7].

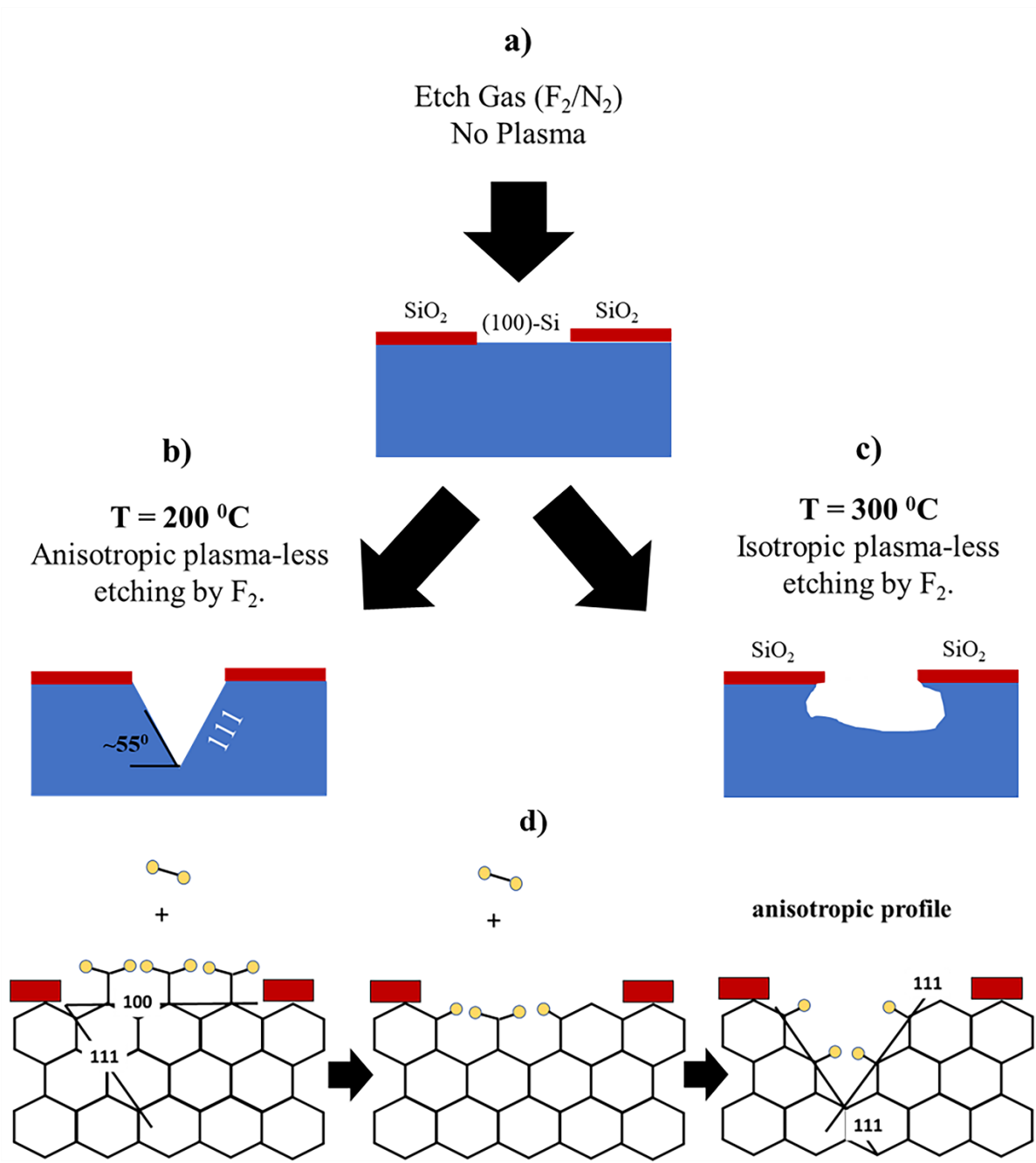

FIG. 1. A schematic representation of ADE of the F-terminated Si surfaces at different substrate temperatures: (a) initial Si(100) surface structure; surface structure after etching by $F_2$ molecules as observed by Kafle et al.[3]: (b) at 200°C which is anisotropic (orientation-dependent); and (c) surface structure after etching at 300°C which becomes more isotropic. (d) Schematic representation of etch profile formation during orientation dependent etching of silicon by $F_2$.

To identify the reasons for the anisotropy we performed quantum chemistry modeling of $F_2$ dissociative chemisorption on silicon surfaces. In our modeling we assume that the etching of silicon by $F_2$ proceeds on the F-terminated surfaces and that the rate of the silicon etching is determined by the rate of $F_2$ chemisorption. The calculated reaction rates of $F_2$ dissociative chemisorption on Si(100) and Si(110) match the rate of silicon etching by $F_2$ gas as measured by Mucha et al. [8] (see Section IID). In addition, our calculations reproduce the experimental trend that $F_2$ gas etches the Si(111) surface more slowly compared to other surface orientations as illustrated in Fig. 1d). The charge density analysis (described in Section IIC) showed that passivating F atoms attract electron density from the uppermost Si atoms and that the amount of positive charge on an individual Si atom increases with the number of bonded passivating F atoms. Si(100) and Si(110) surfaces are passivated by a larger amount of F atoms per surface Si atom compared to Si(111) (Section IIA). $F_2$ molecules approaching the surface become negatively charged and, as a result, are more strongly attracted to the Si(100) and Si(110) surfaces than the Si(111) surface. Analysis shows that the etching limiting step is $F_2$ dissociative chemisorption on silicon and their activation barriers ($E_a$) for Si(100) and Si(110) are approximately equal, $E_a$ (Si(100)) $\approx E_a$ (Si(110), and are small compared to $E_a$ (Si(111)) (Section IV).

Additionally, we performed classical molecular dynamics simulation of etching. It was determined that the REBO potential which was typically used for this chemistry reasonably represents chemisorption on Si(111) but did not accurately represent the reaction path for $F_2$ dissociative chemisorption on Si(100) and Si(110) surfaces, overestimating the required activation energy. As a result, the anisotropic orientation-dependent etching process which creates cone-like nanostructures cannot be simulated using this potential without further reparameterization. The results can be found in Section III.

## II. QUANTUM CHEMISTRY MODELLING
### A. *Structures of silicon surfaces and $F_2$ chemisorption*

As well known the bulk silicon crystallizes in a diamond cubic crystal structure in which each atom has four nearest neighbors forming the sp 3 hybridized tetrahedral structure. The structures of the fluorinated Si(100), Si(110), and Si(111) surfaces and the top 6 silicon atoms ,so-called $Si_6$ chairs, on these surfaces are shown in Fig. 2. The reaction path of $F_2$ dissociative chemisorption passes via the transition state, where a $F_2$ molecule adsorbed on the $Si_6$ chair forms a reaction center with attacked Si-Si. In other words, only two atoms in the $Si_6$ chair are partaking in the reaction, and the stoichiometry of the reaction centers is $F_2..Si_2F_n$. The reaction centers have different "n" depending on the surface orientation. Namely, there is a $F_2..Si_2F_2$ reaction center on the Si(100) and Si(110) surfaces and a $F_2..Si_2F_1$ reaction center on the Si(111) surface as illustrated in Fig. 2. Note that the reaction centers shown in Fig. 3 are coupled with possible reaction channels A and B: one F atom of the $F_2$ molecule bounces off the surface yielding a free F atom or both F atoms are adsorbed without formation of free atoms. Both channels are illustrated in Figure 3 for all $Si_6$ chair configurations.

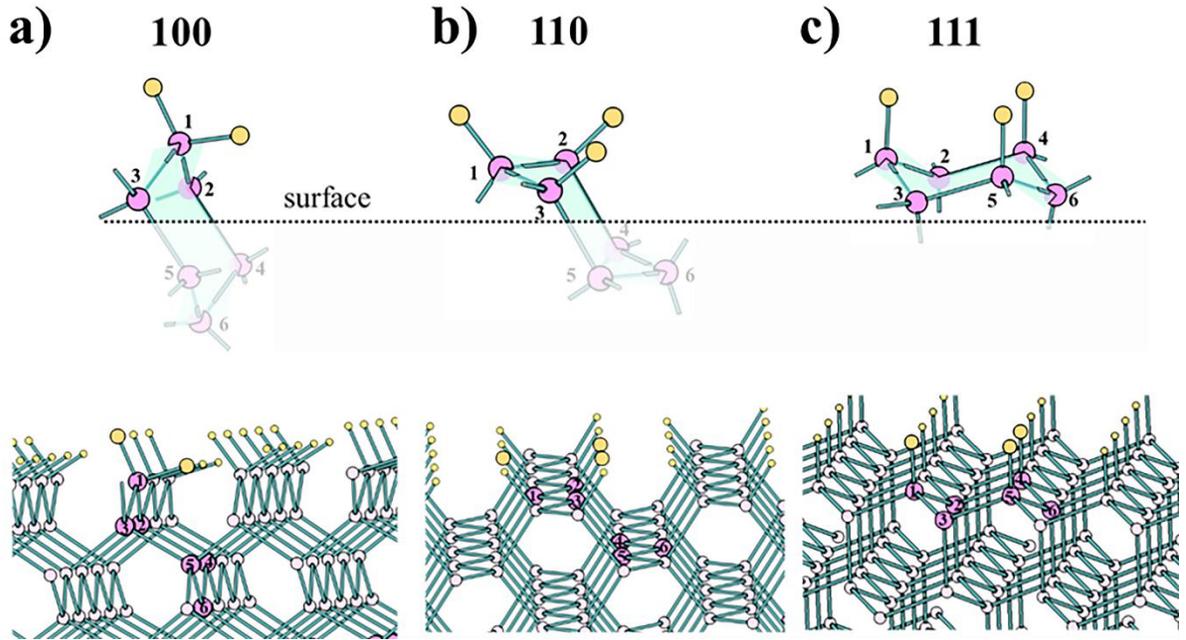

FIG. 2. Fluorinated Si$_6$ chair on 100 (a), 110 (b), and 111 (c) oriented silicon surfaces. Fluorine and silicon atoms are shown in yellow and pink.

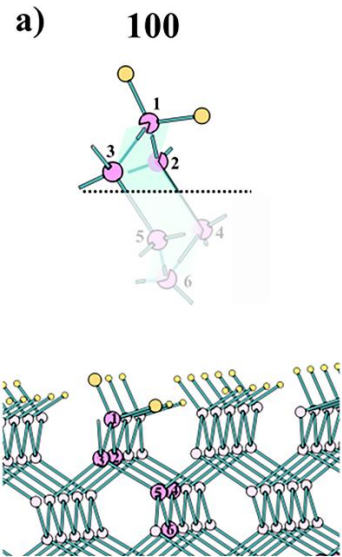
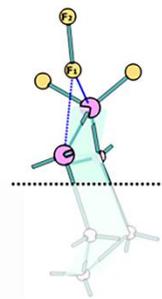
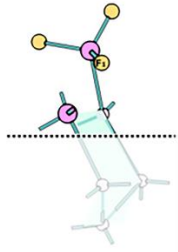
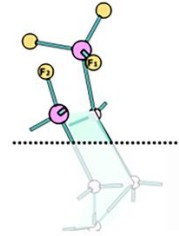
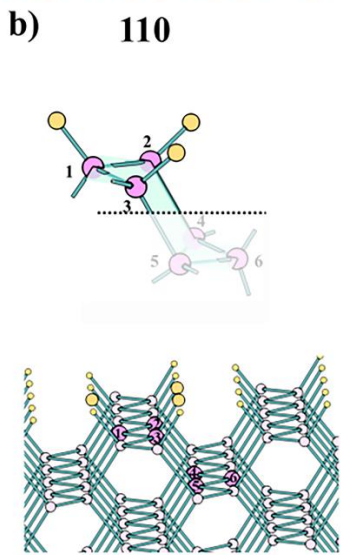
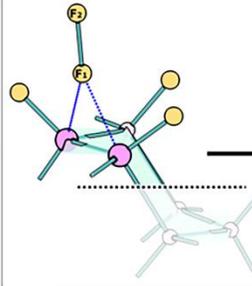
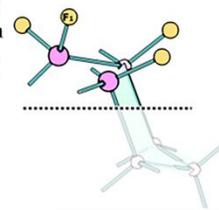
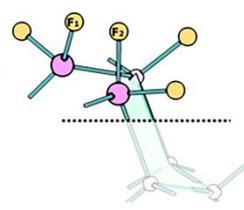
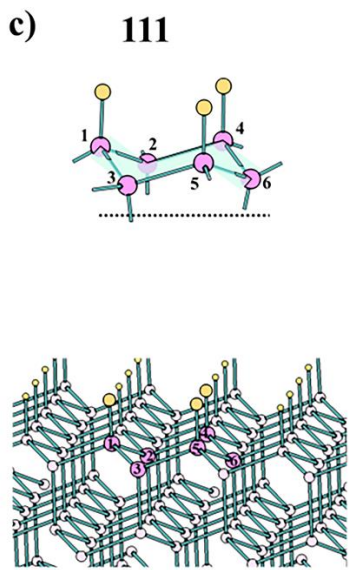
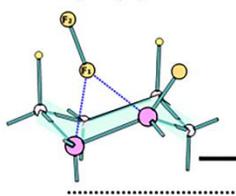
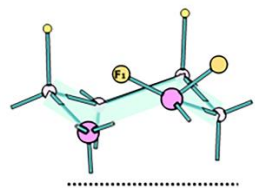
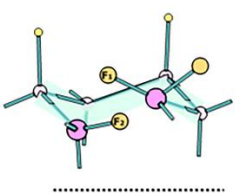

FIG. 3. Dissociative chemisorption of $F_2$ molecules on fluorinated $Si_6$ chairs on 100 (a), 110 (b), and 111 (c) oriented surfaces via $F_2..Si_2F_n$ transition state and leading to single F atom abstraction or two F atoms adsorption. Fluorine and silicon atoms are shown in yellow and pink.

It should be noted that the Si(100) surface can undergo reconstruction to minimize the energy of the top layer [9]. Therefore, in Section IIC we also consider $F_2$ dissociative chemisorption on F-terminated Si(100) reconstructed surfaces as well. We showed that the reconstructed F-terminated Si(100) surface is more reactive (less stable) with $F_2$ molecules than the unreconstructed Si(100) surface. Thus, we assumed that the etching of silicon in the (100) direction is not accompanied by reconstruction and proceeds on unreconstructed Si(100) surface.

## B.  Mechanism of silicon etching by $F_2$

As mentioned in Section IIA, the chemisorption of $F_2$ proceeds via two competitive reaction channels known as chemisorption with single atom abstraction (channel A) and with two atom adsorption (channel B). Channel A requires four elementary steps to etch one silicon atom, while channel B requires only two, as shown in Fig. 4. The first step is the same for both reaction channels A and B. The divergence between channels A and B appears when the TS goes down to the reagent valley, resulting in two different products (see Section IIC). The importance of channel A is that free F atoms are released, which should affect the etch profile. These highly reactive free F atoms can also etch the silicon surface, but rather than the anisotropic, orientation-dependent etch which is possible with molecular fluorine, atomic F etches Si(100), Si(110), and Si(111) with the same rate. Thus, chemisorption which proceeds via channel A causes the etch profile to become more isotropic.

Thus, we performed quantum chemistry modeling and found TSs of the first step in $F_2$ chemisorption on Si(100), Si(110), and Si(111) surfaces. Other steps and the evolution towards products in channels A and B are not considered. However, the calculated reaction barriers of Step 1 match the measured activation energies for silicon etching by molecular fluorine (see Section IV). This agreement confirms our assumption that Step 1 is the rate-determining step of the etching process.

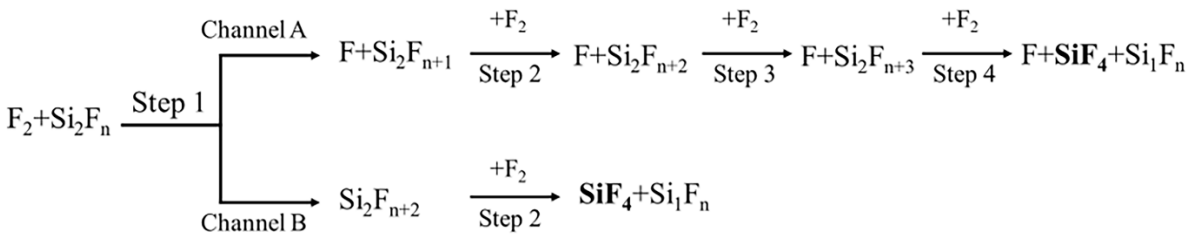

FIG. 4. Reaction mechanisms of $Si_2F_n$ silicon surface etching ($SiF_4$ desorption) by $F_2$ molecules via channels A and B. The first step is shared by both channels, which have the same TS.

## C. Computational details

Quantum chemistry modelling was performed using broken symmetry orbitals calculated by the unrestricted u-B3LYP DFT functional in the Gaussian 16 [10] software package. The Lanl2dz basis set and pseudopotential for Si atoms were used in our simulations. All structures were visualized by Chemcraft software [11].

A $Si_{32}F_{32}$ cluster was used to mimic the structure of fluorinated Si surfaces. The free bonds of external Si atoms were closed by fluorine atoms. The reaction center geometries of four transition states (TSs) corresponding to the reactions of $F_2$ dissociative chemisorption on a Si(100) surface with and without surface reconstruction, on a Si(110) surface, and on a Si(111) surface are shown in Fig. 5. In all transition states considered here the F-F bond is perpendicular to the Si-Si bonds. On the reaction path after the transition state, an intermediate bi-radical structure is observed where singlet and triplet states are overlapping, as shown in Fig. 6. Natural population analysis was performed to calculate atomic charges. The charges on the $Si_2F_n$ reaction centers on the $Si_{32}F_{32}$ cluster before the reaction and in TS can be found in Fig.7. (The $Si_6$ chair of the $Si_{32}F_{32}$ cluster reacting with an $F_2$ molecule is shown in the displayed structure, while other atoms were omitted to simplify the picture). The charge distribution is quite complex; the total charge of the reaction center is not zero, but the total charge of whole $F_2..Si_{32}F_{32}$ molecular system is zero. As shown in Fig. 7, the charge on the Si atoms of $F_2..Si_2F_n$ reaction centers increases slightly in as $F_2$ molecules approach the Si-Si bond due to attraction of electron density by the $F_2$ molecules. However, the majority of positive charge on these Si atoms is due to electron displacement to F substitutes that form part of the reaction center, not due to electron displacement to $F_2$ molecules. Thus, the reactivity of the surface is determined by the surface stoichiometry, which in turn determines the surface charges and the reaction barrier for $F_2$ dissociation.

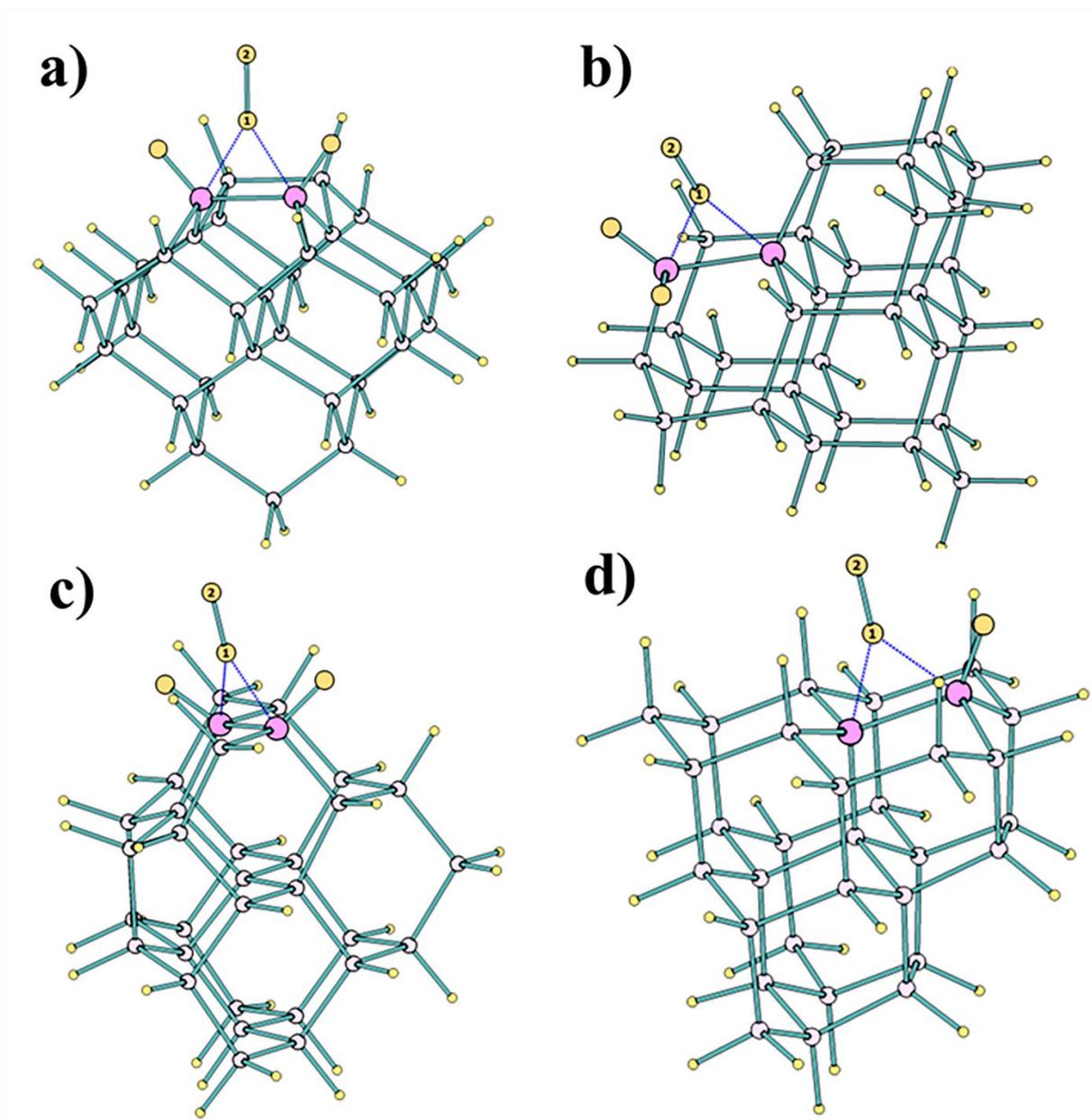

FIG. 5. Geometries of transition states of F$_2$ dissociative chemisorption on reconstructed 100 (a), unreconstructed 100 (b), 110 (c), and 111 (d) surfaces of the Si$_{32}$F$_{32}$ cluster. Fluorine and silicon atoms are shown in yellow and pink.

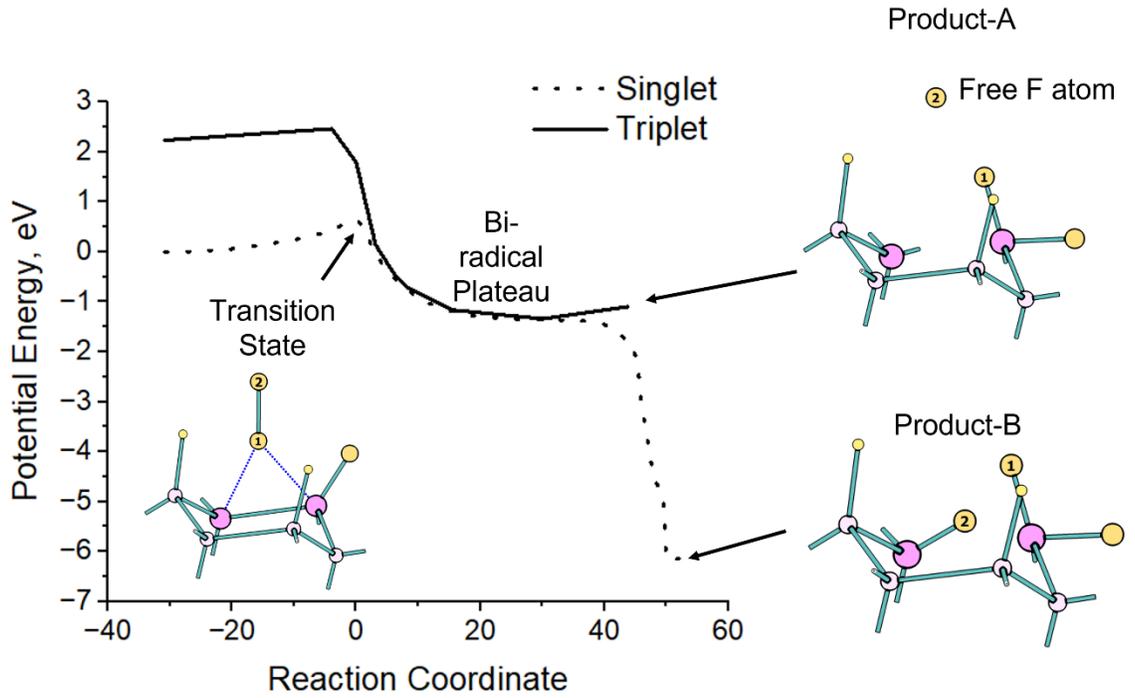

FIG. 6. Reaction pathway of $F_2$ dissociative chemisorption on a 111-oriented surface of the $Si_{32}F_{32}$ cluster. Fluorine and silicon atoms are shown in yellow and pink.

Pre-exponential factors in the Arrhenius equations were calculated using the Eyring equation [12]:

$$A = \frac{k_B T}{h} \frac{Z_{vib}^{TS}}{Z_{vib}^{Si32F32} Z_{tot}^{F2}} \qquad (1)$$

where $k_B$ and $h$ are Boltzmann's and Planck's constants, $Z_{vib}^{TS}$ and $Z_{vib}^{Si32F32}$ are vibrational partition functions of the TS geometry and the isolated $Si_{32}F_{32}$ cluster, and $Z_{tot}^{F2}$ is the total partition function of the isolated $F_2$ molecule. The partition functions in eq(1) were calculated using Gaussian 16 software. The pre-exponential factor $A$ in eq(1) is a function of temperature; therefore it was fitted using the standard form in a temperature range from 298.15 to 1000K:

$$A = A'(T/298.15)^{n1} \qquad (2)$$

The rate of the surface reaction can be described either as a reaction probability or in standard Arrhenius form:

$$\frac{1}{4}\gamma \sqrt{\frac{8k_B T}{\pi m}} C_{F2} = \frac{k_B T}{h} \frac{Z_{vib}^{TS}}{Z_{vib}^{Si32F32} Z_{tot}^{F2}} exp(-\frac{E_a}{k_B T}) \rho_s C_{F2} \qquad (3)$$

where $m$ and $C_{F2}$ are the mass and density of $F_2$, $E_a$ is the activation energy of the reaction and $\rho_s$ is the surface site density. Thus, the probability $\gamma$ of $F_2$ dissociative chemisorption can be calculated using the next equation:

$$\gamma(T) = \gamma_0(T)\sqrt{\frac{T}{T_0}} \exp\left(-\frac{Ea}{k_BT}\right) = \gamma_0'\left(\frac{T}{T_0}\right)^{n2} \exp\left(-\frac{Ea}{k_BT}\right) \qquad (4)$$

where $\gamma_0(T) = \frac{\sqrt{2\pi m k_B}}{h} \frac{Z_{vib}^{TS}}{Z_{vib}^{Si32F32} Z_{tot}^{F2}} \rho_s \sqrt{T_0}$ depends on temperature, because the partition functions are functions of temperature. Hereinafter we will assume that $T_0 = 298.15$ K. $\gamma_0'\left(\frac{T}{T_0}\right)^{n2}$ is a function, where $\gamma_0'$ and $n2$ parameters were chosen to fit calculated $\gamma_0(T)\sqrt{\frac{T}{T_0}}$ values at a temperature range from 298.15 to 1000K. The values of $E_a$, A, $\gamma_0'$, $n1$ and $n2$ can be found in Table 1, where these parameters are substituted into the expression of rate constants and probabilities:

$$k(T) = A_0 \left(\frac{T}{298.15}\right)^{n1} \exp\left(-\frac{Ea}{k_BT}\right) \qquad (5a)$$

$$\gamma(T) = \gamma_0' \left(\frac{T}{298.15}\right)^{n2} \exp\left(-\frac{Ea}{k_BT}\right) \qquad (5b)$$

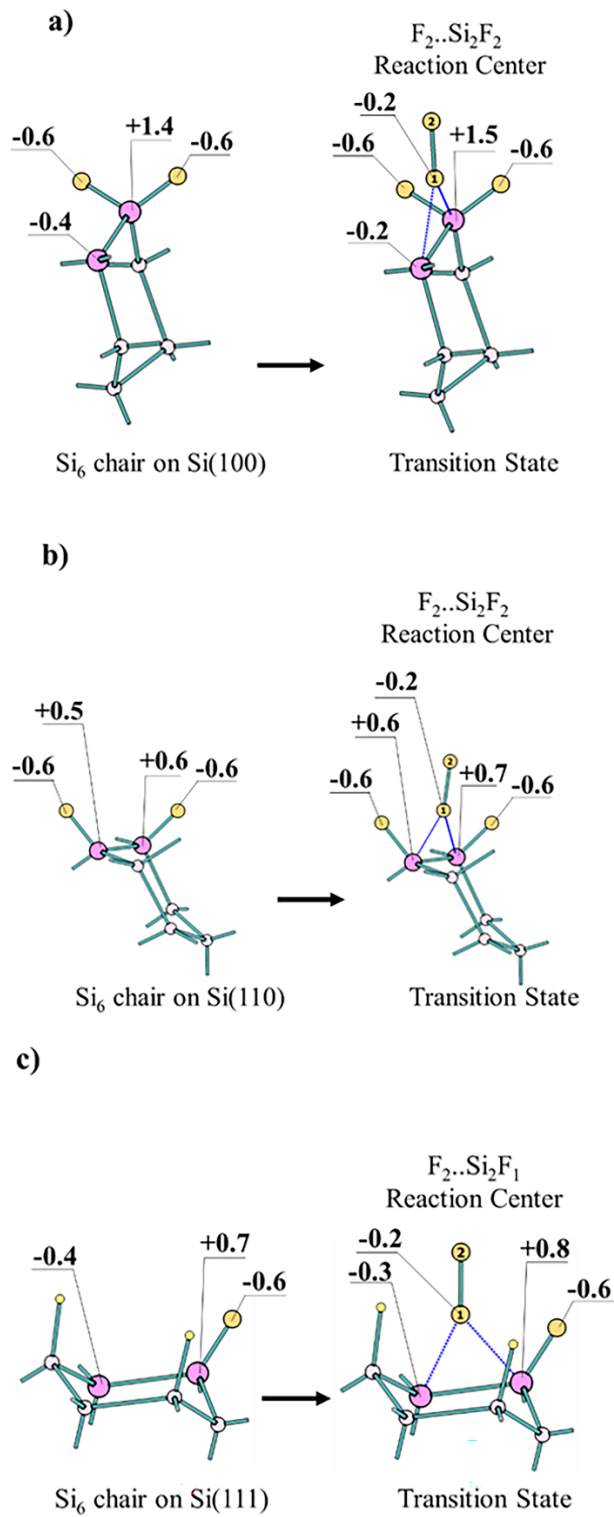

FIG. 7. Charge distribution on the Si$_2$F$_n$ reaction centers on 100 (a), 110 (b), and 111 (c) surfaces of Si$_{32}$F$_{32}$ cluster before the reaction with F$_2$ (Step 1) and in TSs. To simplify the picture the atoms of $Si_6$ chair are presented only, while other atoms were omitted. Fluorine and silicon atoms are shown in yellow and pink.

## D. DFT validation

The presence of the biradical structure in Fig. 6 is consistent with dissociative chemisorption via single atom abstraction as proposed by Tate et al. [13,14], in which only one atom of $F_2$ forms a Si-F bond while the other F atom desorbs into gas phase (channel A in Fig. 4 and product A in Fig. 6). Channel B in Fig. 4 and product B in Fig. 6 corresponds to the reaction channel of dissociative chemisorption, where both atoms of the $F_2$ molecule adsorb on the silicon surface resulting in two Si-F bonds. Note that molecular dynamic simulations by Carter et al. [15] showed that both reaction channels yielding product A and product B are present and they are the dominant reaction channels on a clean, unpassivated Si(100) surface. We expect that the reaction channel with single atom abstraction decreases the selectivity of orientation-dependent etching because one of the F atoms of the $F_2$ molecule (which initially desorbs into the gas-phase) is re-deposited on the substrate and non-selectively etches the Si(100), Si(110) and Si(111) surfaces.

As mentioned in the introduction, Mucha et al. [8] measured the rate of silicon etching by $F_2$ gas. Unfortunately, they did not mention the orientation of the silicon surface exposed to the reactants. Nevertheless, we can compare our etch rate ($R$) calculated using equation (6) (Section IIA) with the measured $R$ by Mucha at al..

$$R = A * \left(\frac{T}{298.15}\right)^n * \exp\left(-\frac{E_a}{RT}\right) * \frac{M(Si)}{p(Si)*N_A} * p_s, \quad (6)$$

$$Etch\ Rate = R * n_{F2}, \quad (7)$$

where $M(Si)$ is the molar mass of silicon, $p(Si)$ is the density of bulk silicon, $N_A$ is Avogadro's number, $p_s$ is the surface site density, $A$ and $E_a$ are the pre-exponential factor and activation energy of $F_2$ dissociative chemisorption on a $Si_{32}F_{32}$ cluster (Section IIA).

The results are presented in Fig. 8, which shows that our calculations give good agreement with experimental data if we assume that the exposed surface in the experiment is the Si(100) facet or the Si(110) facet. (We arbitrarily chose the density of $n(F_2)$ to be equivalent to 1 Torr partial pressure in order to present the result in convenient units, monolayers per second.) Note that the effect of atomic F released by the single atom abstraction mechanism is not considered in our calculation of the etch rate in eq.(7) and Fig. 8.

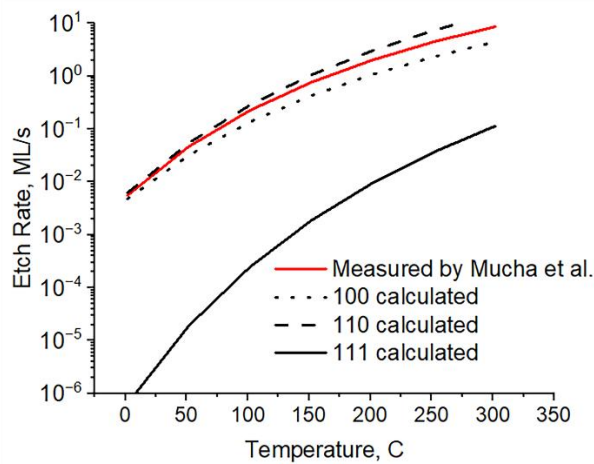

FIG. 8. Calculated etch rate of unreconstructed Si(100), Si(110) and Si(111) surfaces by 1 Torr $F_2$ in monolayers per second (ML/s), compared to measured etch rate by Mucha et al [8] in solid red.

Pullman et al. [16] experimentally studied dissociative chemisorption of $F_2$ on a Si(100)(2×1) surface saturated with a single monolayer of fluorine by exposing the fluorine-saturated surface to supersonic $F_2$ beams of variable energy. They found that no reaction occurs below 0.16 eV of incident energy. Note that this experiment aimed to study the reaction of $F_2$ dissociative chemisorption without subsequent etching. Therefore, we assume that the studied reaction proceeded on initially relaxed Si(100) with surface reconstruction. Indeed, our calculated reaction barrier for $F_2$ dissociative chemisorption on the reconstructed Si(100) surface (0.13 eV) is close to the measured threshold value of $F_2$ incident energy (0.16 eV). In other words, the measured threshold value of $F_2$ incident energy is determined by the barrier of $F_2$ dissociative chemisorption on the reconstructed surfaces.

The reaction of molecular fluorine with a Si(111) surface was studied by Tatsumi and Hiroi [17] in an ultra-high vacuum reactor. They found that the effective activation energy for the surface reactions of molecular fluorine leading to Si(111) surface etching at temperatures below 580 $^0$C was 0.61 eV. According to our calculations, the reaction barrier for $F_2$ dissociative chemisorption on Si(111) is 0.57 eV.

## III. Classical Molecular Dynamics

We also performed classical molecular dynamics (MD) simulations to deepen our understanding of the crystallographic orientation-dependent etching of Si by $F_2$. In these simulations, Si-F interactions are described using a reactive empirical bond order (REBO) potential. These types of force fields are commonly used to simulate covalently bonded materials (such as Si), as well as their reactions with a variety of gaseous species. We use the REBO potential parameterized by Humbird and Graves[18].

In order to utilize the MD code in an efficient manner, we simulate the etching process by a series of "impact simulations". These simulations consist of a semi-infinite Si slab with a vacuum space above the slab in the *z* direction. The Si slab consists of Si atoms in a diamond-cubic lattice. Periodic boundary conditions are imposed in the *x* and *y* directions to make the slab semi-infinite. The bottom two layers are fixed to prevent drift of the slab due to incoming $F_2$ impacts. During an "impact simulation" a $F_2$ molecule is placed in a random position in the vacuum space above the Si slab. The height of the molecule is chosen such that it is just outside the interaction distance of the Si slab. The velocity components of the $F_2$ are randomly sampled from the Maxwell-Boltzmann distribution at 300 K. We only consider molecules with a negative *z* component of velocity, so the molecule will impact the Si slab. Next, a microcanonical ensemble (constant number of atoms, constant volume, and constant total energy) MD simulation is run lasting approximately 2 ps. After the trajectory is complete, species that have been etched or sputtered are deleted from the simulation. In addition, species that are weakly bound to the surface are also deleted. After this, a Berendsen thermostat[19] is applied to the simulation cell to remove excess energy introduced by the incoming species, and to maintain the temperature of the Si slab. More information on the thermostat procedure and product deletion routines can be found in previous work[20,21]. This procedure is repeated until the desired fluence of $F_2$ is reached. We consider Si slabs with different crystal facets facing the vacuum space to study the different etching behavior of the facets. The surface area for the Si slab with the (100) surface exposed (both unreconstructed and reconstructed) is about 472.5 $Å^2$. For the slab with the (110) surface exposed, the surface area is about 626.4 $Å^2$. The surface area for the Si slab with the (111) surface exposed is roughly 613.8 $Å^2$. The depth of each cell is large enough such that no $F_2$ molecules will interact with the fixed layer. The lattice constant in all cells corresponds to diamond cubic Si at 300 K. Even though we consider temperatures higher than 300 K, the lattice constant remains the same in our simulations. In reality, the crystal will expand as temperature is raised. However, because the thermal expansion of Si is very small (the lattice constant increases by about 0.5% from 298.5 K to 1513.2 K[22]) we ignore this expansion in our simulations.

Our classical MD simulations did not show silicon etching at temperatures below 900 K. ($F_2$ bounced off of the silicon surfaces every time.) In order to explain the absence of etching we performed intrinsic reaction coordinate (IRC) DFT calculations of $F_2$ dissociative chemisorption on $Si_{32}$ clusters for all TSs shown in Fig. 3. IRC calculations follow the steepest descent path from the TS to the reagent and product valleys. The same reaction pathways were recalculated using the REBO classical MD potential. As shown in Fig. 9(a), (b) and (c), very high barriers appear on the (100) and (110) reactions paths recalculated by the classical MD potential which do not appear on the DFT-calculated reaction paths. The (111) reaction path, on the other hand, is quite accurately reproduced by the REBO potential.

The lack of etching observed in molecular dynamics simulations at temperatures below 900 K can be explained by the high reaction barriers. The reactions are very slow below 900 K and as a result cannot be simulated by classical MD in a reasonable amount of time. In the case of Si(100) and Si(110) surface etching, the barriers are overestimated and additional fitting of the potential would be required for Si + $F_2$ chemistry in order to correct this. Without reparameterization, the MD potential would be unable to simulate the anisotropic etching described earlier in the paper. The reaction barrier for $F_2$ dissociative chemisorption on Si(111) calculated by the REBO potential is very close to the DFT value (0.61 eV vs. 0.57 eV). However, the inverse reaction rate (timescale of the reaction) is still higher than 1 ns at temperatures below 900 K (see Fig. 6(d)). The etching probabilities calculated by classical MD and the probabilities of $F_2$ dissociative chemisorption calculated by transition state theory can be found in Fig. 10.

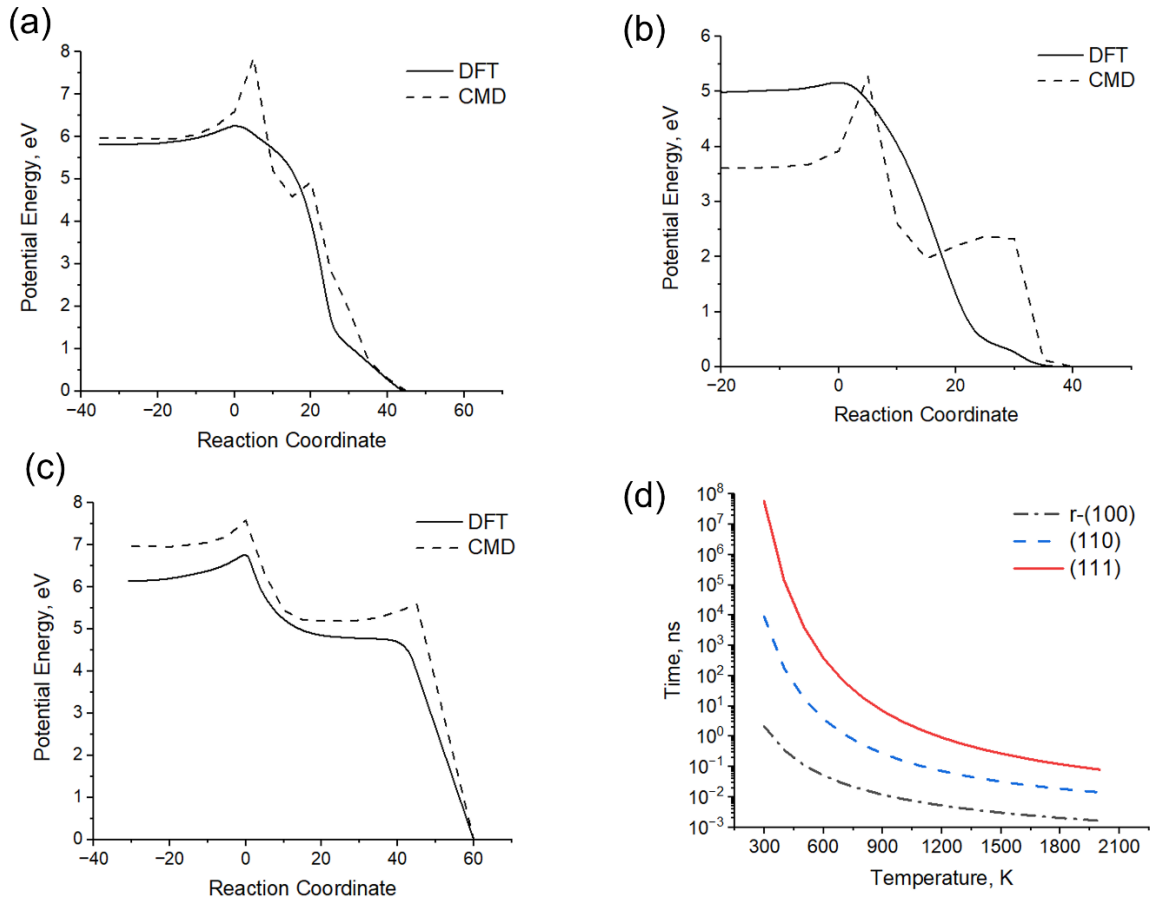

FIG. 9. Reaction path of $F_2$ dissociative chemisorption on reconstructed (100) (a), (110) (b) and (111) (c) surfaces of the $Si_{32}$ cluster calculated by DFT and recalculated with the REBO classical MD (CMD) potential. Inverse reaction rates were calculated by DFT (d).

## IV. RESULTS AND DISSCUTION

As our DFT calculations show, the dissociative chemisorption of $F_2$ on Si(111) proceeds over a significantly higher barrier than dissociative chemisorption on Si(100) and Si(110) surfaces. This results in a lower reaction probability (Fig. 10) and hence a lower etch rate for the Si(111) surface. We also demonstrate that etching of the (100) surface with reconstruction is faster than etching on the (100) surface without reconstruction. We assume that the removal of surface layers by fluorine proceeds without reconstruction. Thus, it is equally probable that the $F_2$ molecule dissociates on unreconstructed (100) and (110) surfaces, so etching in the (100) and (110) directions proceeds at the same rate. The low etch rate of the Si(111) surface by comparison explains the orientation-dependent etching which was used by Kafle et al. [3–6] for texturing silicon surfaces during black silicon production.

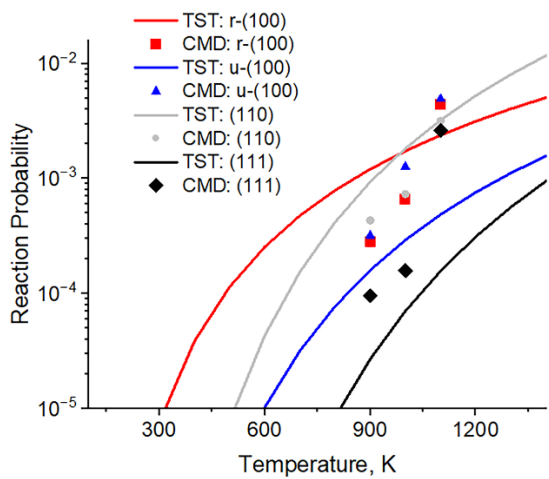

FIG. 10. The probabilities of $F_2$ dissociative chemisorption on a $Si_{32}F_{32}$ cluster calculated by transition state theory (TST) under DFT approach and probabilities of the silicon slab etching calculated by classical MD (CMD) for (100), (110) and (111) silicon surfaces.

We attribute the high etch rate of Si(100) and Si(110) to the fact that these surfaces are passivated by a larger number of F atoms per surface Si atom compared to Si(111) surfaces. The electron density is displaced from Si atoms to the more electronegative passivating F atoms, resulting in partial positive charge on Si atoms (Fig. 11). (This effect is known as the inductive effect in organic chemistry). The amount of positive charge on a surface increases with "n", and as a result Si(100) and Si(110) surfaces accumulate larger amounts of positive charge relative to the Si(111) surface. Thus, the reaction barrier for dissociative chemisorption of molecular fluorine is smaller for Si(100) and Si(110) surfaces compared to Si(111) surfaces (Fig. 11b). An $F_2$ molecule approaching the silicon surface becomes negatively charged, attracting electron density from surface Si atoms. As a result, $F_2^{\delta-}$ is more strongly attracted to the Si(100) and Si(110) surfaces than to the Si(111). This leads to the observed fast etching in the (100) and (110) directions and slow etching in the (111) direction.

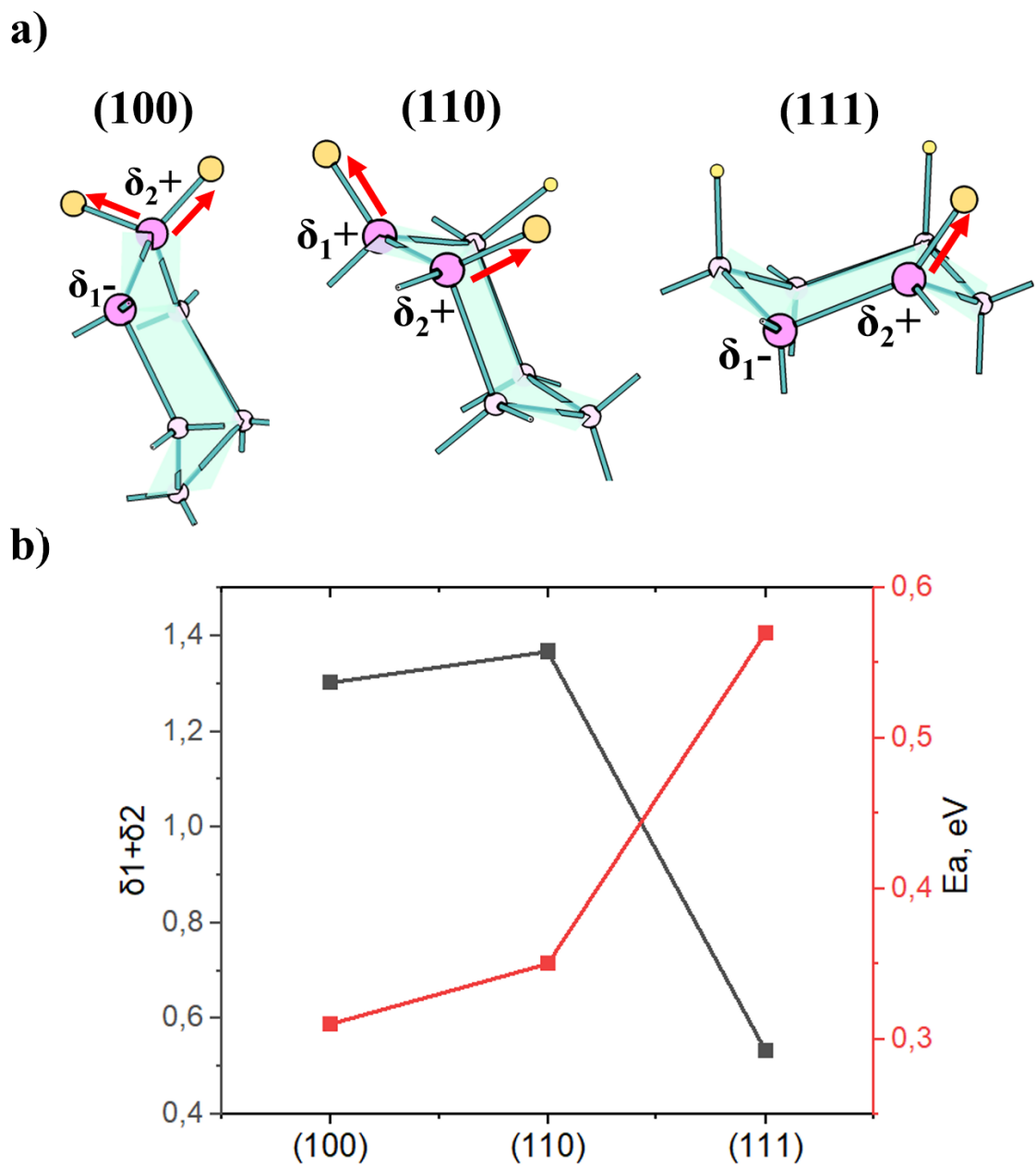

FIG. 11. The electron displacement (red arrow) leading to δ1 + δ2 charges on the $Si_2F_n$ reaction centers caused by electronegative F substitutes on F-terminated Si(100), Si(110), and Si(111) surfaces (a). (Fluorine and silicon atoms are shown in yellow and pink). The calculated δ1 + δ2 charge on Si-Si dimer of the $Si_2F_n$ reaction center on Si(100), Si(110), and Si(111) surfaces and the barriers ($E_a$) of $F_2$ dissociative chemisorption on these dimers (b).

Our simulations show that the activation barriers of $F_2$ dissociative chemisorption are: 0.31 eV on Si(100), 0.35 eV on Si(110) and 0.57 eV on Si(111). Finally, we collect calculated parameters for rate constants and probabilities of $F_2$ dissociative chemisorption on silicon surfaces in Table 1.

TABLE 1. Rate constants and probabilities of $F_2$ dissociative chemisorption on the (100), (110) and (111) surfaces of a $Si_{32}F_{32}$ cluster. The rate constants and the probabilities are expressed in Arrhenius form according to equations 5a and 5b. $E_a^{exp}$ is the experimental value of the activation energy. r-100 and u-100 indicate the (100) surface with reconstruction and without reconstruction, respectively.

| Surface | A', m3/s | n1 | $\gamma_0$` | n2 | $E_a$,eV | $E_a^{exp}$,eV |
|---------|----------|------|--------|------|------|---------|
| r-100   | 8.82·10⁻²¹ | 2.20 | 0.0011 | 1.60 | 0.13 | 0.16[16] |
| u-100   | 1.23·10⁻²⁰ | 2.20 | 0.0016 | 1.70 | 0.31 | 0.39[8]  |
| 110     | 7.52·10⁻²⁰ | 2.50 | 0.0096 | 1.96 | 0.35 | -       |
| 111     | 3.90·10⁻²⁰ | 2.50 | 0.0050 | 1.96 | 0.57 | 0.61[17] |

# V. CONCLUSION

Orientation-dependent etching of silicon via molecular $F_2$ has been studied using quantum chemistry methods. Activation barriers, reaction rates, and probabilities for $F_2$ dissociative chemisorption on (100), (110), and (111) surfaces were calculated. It was shown that the reaction barrier of $F_2$ dissociative chemisorption is a function of the surface stoichiometry, which in turn determines the amount of positive charge on surface Si atoms. As a result, the etching of Si(111) is significantly slower than (100) and (110) surfaces. Our simulation results are consistent with previously published experimental data and explain the mechanism of orientation-dependent etching of Si by $F_2$ gas.

Additionally, we describe two reaction channels of $F_2$ dissociative chemisorption. We expect that the desorption of F atoms during the $F_2$ dissociation is significant during the etching process. This is consistent with dissociative chemisorption via single atom abstraction, which was proposed previously. We expect that the reaction channel via single atom abstraction causes non-orientation-dependent etching with isotropic profile.

We also used molecular dynamics simulations to investigate the suitability of a REBO potential for modelling silicon etching via molecular $F_2$. While the REBO potential reasonably described $F_2$ dissociative chemisorption on Si(111), it predicts unrealistically high reaction barriers for dissociative chemisorption on Si(110) and Si(100). This results in very slow etching rates and an inability to reproduce the anisotropic etching mechanism studied in this work. In the future, molecular dynamics modelling of this process will require reparametrized potentials that accurately describe orientation-dependent reaction barriers.

## ACKNOWLEDGMENT

The research described in this paper was conducted under the Laboratory Directed Research and Development (LDRD) Program at Princeton Plasma Physics Laboratory, a national laboratory operated by Princeton University for the U.S. Department of Energy under Prime Contract No. DE-AC02-09CH11466. The United States Government retains a non-exclusive, paid-up, irrevocable, world-wide license to publish or reproduce the published form of this manuscript, or allow others to do so, for United States Government purposes.

In addition, this research used computing resources on the Princeton University Adroit Cluster and Stellar cluster. The authors gratefully acknowledge discussions about the classical MD code with David Humbird (DWH Process Consulting).

## AUTHOR DECLARATIONS

**Conflicts of Interest**

The authors have no conflicts to disclose.

## DATA AVAILABILITY

The data that support the findings of this study are available within the article as well as from the corresponding author upon reasonable request.